\documentclass[fleqn,10pt]{olplainarticle}

\usepackage{siunitx}
\usepackage{lineno}
\usepackage{todonotes}
\newcommand{\tsup}{\textsuperscript}
\newcommand{\tsub}{\textsubscript}

\title{High-speed detection of 1550 nm single photons with superconducting nanowire detectors}

\author[1,*]{Ioana Craiciu}
\author[1]{Boris Korzh}
\author[1]{Andrew D. Beyer}
\author[1,2]{Andrew Mueller}
\author[1]{Jason P. Allmaras}
\author[2]{Lautaro Narvaez}
\author[2]{Maria Spiropulu}
\author[1]{Bruce Bumble}
\author[3]{Thomas Lehner}
\author[1]{Emma E. Wollman}
\author[1]{Matthew D. Shaw}

\affil[1]{Jet Propulsion Laboratory, California Institute of Technology, 4800 Oak Grove Dr, Pasadena, California 91109, USA}
\affil[2]{Division of Physics, Mathematics and Astronomy, California Institute of Technology, 1200 E California Blvd, Pasadena, California 91125, USA}
\affil[3]{Dotfast Consulting, Untere Sparchen 16, 6330 Kufstein, Austria}

\affil[*]{Corresponding author: ioana.craiciu@jpl.nasa.gov}

\begin{abstract}
Superconducting nanowire single photon detectors are a key technology for quantum information and science due to their high efficiency, low timing jitter, and low dark counts. In this work, we present a detector for single 1550 nm photons with up to 78\% detection efficiency, timing jitter below 50 ps FWHM, 158 counts/s dark count rate – as well as a world-leading maximum count rate of 1.5 giga-counts/s at 3 dB compression. The PEACOQ detector (Performance-Enhanced Array for Counting Optical Quanta) comprises a linear array of 32 straight superconducting niobium nitride nanowires which span the mode of an optical fiber. This design supports high count rates with minimal penalties for detection efficiency and timing jitter. We show how these trade-offs can be mitigated by implementing independent read-out for each nanowire and by using a temporal walk correction technique to reduce count-rate dependent timing jitter. These detectors make quantum communication practical on a 10 GHz clock.\\
\textcopyright\, \the\year{}. All rights reserved.
\end{abstract}

\begin{document}

\maketitle

\section{Introduction}

Superconducting nanowire single-photon detectors (SNSPDs) can detect single photons from mid-infrared \cite{Verma2021,Colangelo2022} to ultraviolet \cite{Wollman2017} wavelengths, with efficiencies up to $>98\%$ \cite{Reddy2020,Chang2021}, dark count rates of $6\times10^{-6}/\mathrm{s}$ \cite{Chiles2022}, timing jitter of \SI{3}{\pico\second} FWHM \cite{Korzh2020}, and count rates of up to 800 mega-counts/s (3 dB compression) \cite{Zhang2019}. They have enabled tests of local realism \cite{Giustina2015}, key demonstrations in quantum information processing \cite{Zhong2020} and quantum communication \cite{Boaron2018PRL}, and optical communication between the earth and the moon \cite{Grein2015}. New applications continue to emerge as their performance improves \cite{Esmaeil2021}. As quantum communication protocols in particular are leveraging the low jitter of SNSPDs to increase clock rates to 10~GHz and beyond~\cite{Takesue2007, Wang2019}, it is increasingly important to measure at high count rates while maintaining low jitter and high efficiency.

When an SNSPD absorbs a photon, a hotspot is created, which disrupts the superconductivity and shunts current from the nanowire to the readout electronics \cite{Goltsman2001}. The maximum rate at which an SNSPD can count photons is limited by the timescale at which the hotspot dissipates. One strategy to increase count rate is to decrease the thermal reset time by choice of superconducting material \cite{Cherednichenko2021} or by using very short nanowires \cite{Vetter2016}. However, these strategies have led to detectors with lower internal detection efficiency as indicated by a lack of saturation in photon count rate with bias current, indicating a trade-off between fast thermal reset dynamics and detection efficiency. Another strategy to overcome the count rate limit is to use several parallel nanowires to measure the signal of interest. In these devices, after one nanowire detects a photon, the other wires are still superconducting and optimally biased for detection. For fiber-coupled modes, efforts have focused on connecting the nanowires in parallel with a single readout line \cite{Korneev2007,Perrenoud2021}. The count rate achievable with this method was limited by electrical cross-talk between wires. High count rates have also be achieved in free-space-coupled detector arrays \cite{Allmaras2017,Zhang2019}, but these arrays used long nanowires which had significant longitudinal geometric timing jitter \cite{Zhao2017}, leading to a jitter of greater than 50~ps full-width-at-half-maximum (FWHM). 

In this work we present the PEACOQ detector, in which 32 short nanowires in a linear array are read out individually. At the cost of increased readout complexity, this nanowire array can detect photons in a single 1550~nm optical mode at a world-leading count rate of 1.5 giga-counts/s (Gcps) at 3 dB compression, along with a system detection efficiency (SDE) of up to 78\% and low timing jitter. We demonstrate a strategy for maintaining a timing jitter below 46 ps FWHM at count rates up to 1 Gcps by building on a recently developed temporal walk correction technique \cite{Mueller2022}. A detector with high efficiency and low timing jitter at high count rates, as well as low dark counts, is required to detect faint optical signals at high clock rates, such as in quantum communication or deep-space optical communication where amplification of optical pulses is not feasible. For example, the PEACOQ can count single-photons at 250 Mcps with an SDE of 70\% and, with realistic improvements to our measurement setup, a jitter of 86~ps full-width-at-1\%-maximum (FW1\%M). This would enable a quantum communication protocol based on a 10 GHz pulsed source with mean photon number per pulse at the detector of $\langle n \rangle=0.025$. 

\section{PEACOQ Overview}\label{sec:device}

\begin{figure}[ht!]
    \centering\includegraphics[width=3.2in]{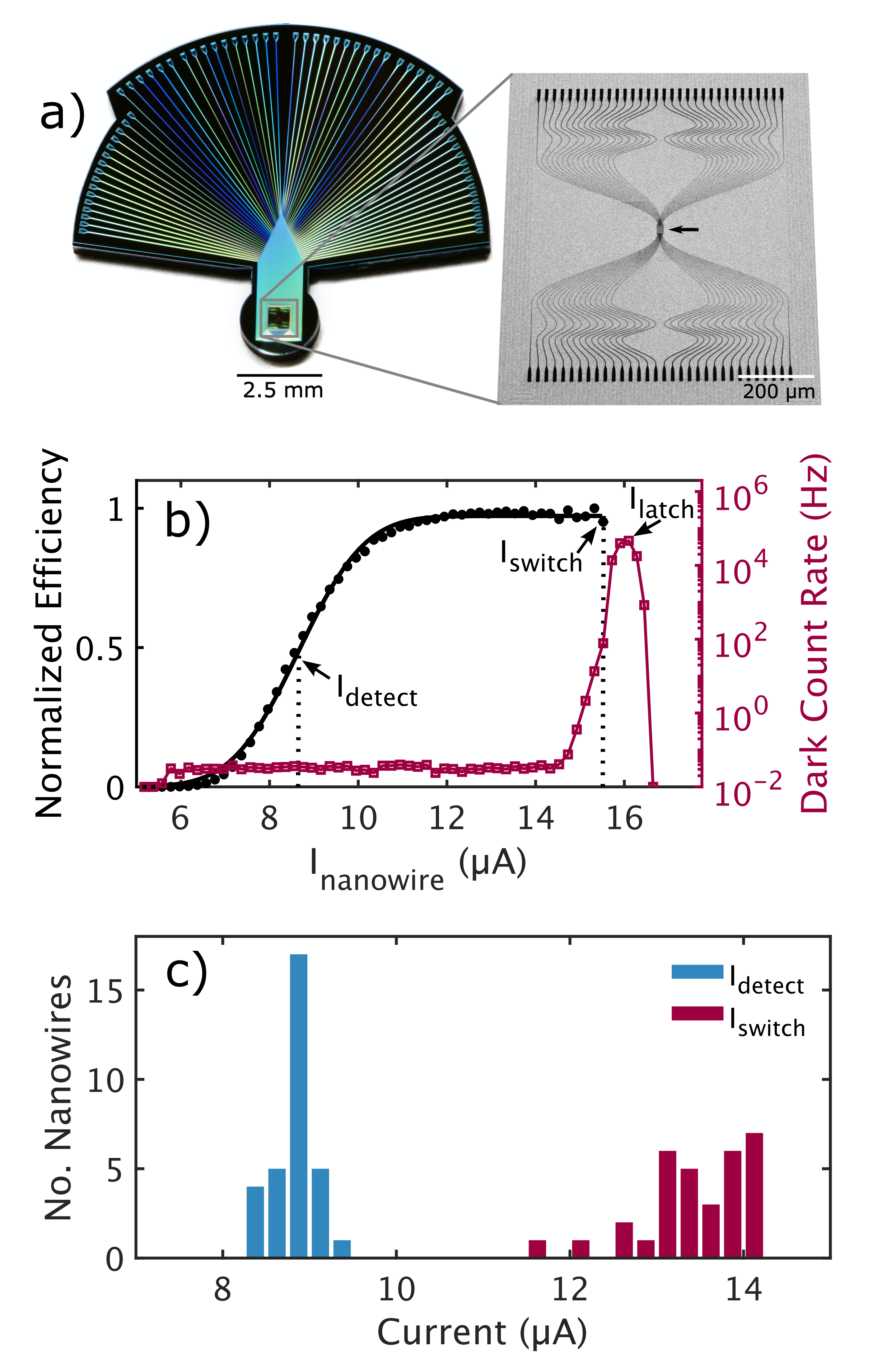}
    \caption{\label{fig:device} a) Photograph of the PEACOQ detector. Inset is a scanning electron micrograph showing the nanowires, microstrips (dark meandered lines), and coplanar waveguides (lighter lines connected to top and bottom of microstrips, also visible  on photograph). The active region, indicated by the arrow, is a linear array of nanowires covering an area of $\SI{13}{\micro\meter}\times\SI{30}{\micro\meter}$. b) Normalized efficiency (black circles, left axis) and dark count rate (red squares connected by solid line, right axis) as a function of bias current in one nanowire. Solid black curve is a fit to the internal detection efficiency equation (see main text). c) $I_\mathrm{detect}$ (blue) and $I_\mathrm{switch}$ (red) for all nanowires in the array. Note that the switching currents measured in this readout configuration are slightly lower than in b) due to different measurement set-ups (see Measurement Setup section)}
\end{figure}

Figure~\ref{fig:device}a shows the PEACOQ detector. The active area of the detector consists of a linear array of 32 parallel niobium nitride (NbN) nanowires on a 400~nm pitch. The wires are 120~nm wide, \SI{30}{\micro\meter}-long, straight sections (no meanders). A two-part transmission line connects each side of each nanowire to bonding pads at the edges of the PEACOQ chip. The first part is an adiabatically tapered microstrip, comprised of a NbN conductor, a dielectric layer (SiO\tsub{2}), and a gold mirror plane below. The second part is a coplanar waveguide (CPW) comprised of the NbN and contact metals (Ti/Nb/Ti/Au/Ti) on SiO\tsub{2} (no gold plane), with a conductor width of \SI{9}{\micro\meter}, and a conductor-to-ground spacing of \SI{3.625}{\micro\meter}. There is a short tapered region connecting the conductors of the microstrip and CPW. The symmetric coupling on both sides of the wire allows for differential readout. In combination with a differential amplifier, differential readout enables cancellation of geometric jitter \cite{Colangelo2021}, common-mode noise rejection, and larger signal to noise, all of which can reduce overall timing jitter. 

The total inductance of the nanowire, which is dominated by the kinetic inductance of the superconducting NbN (nanowire and waveguides), is a key design parameter that affects the maximum count rate and timing jitter. The microstrips serve the purpose of increasing the kinetic inductance of the nanowires, to prevent latching. Short nanowires are used in the design to minimize longitudinal geometric jitter \cite{Korzh2020}, but the kinetic inductance of the \SI{30}{\micro\meter}-long nanowire of $\approx50$ nH is too low, and would lead to latching of the device at low bias currents \cite{Kerman2009}. Each microstrip has a kinetic inductance of $\approx315$~nH each, controlled by varying the width profile along the taper, with $L_k=L_\mathrm{sq}\int_0^\ell\frac{1}{w(s)} d\mathrm{s}$, where $L_\mathrm{sq}=194$~pH is the sheet inductance, $w(s)$ is the width along the microstrip (which varies between 120~nm and \SI{1.5}{\micro\meter}), and $\ell$ is the arc length of the microstrip. The microstrips are meandered to allow them to be roughly equal in length, leading to equal inductance across the 32 nanowires in the array. The sheet inductance is estimated from the reset time of the device $\tau_\mathrm{reset}=\frac{L_k}{R_L}$, which is measured in the Maximum Count Rate section. Device fabrication is described in the Supplemental Materials. 

Figure~\ref{fig:device}a also shows the overall shape of the detector chip. The circular area at the bottom in the picture is 2.5~mm in diameter, designed to align with a standard FC/PC fiber ferrule. Self-alignment between the optical mode and the detector active area \cite{Miller2011} is enabled by a custom designed ceramic sleeve with a wider opening (1.9 mm) that fits around both ferrule and detector.

Figure~\ref{fig:device}b shows the normalized photon count rate (PCR) versus bias current for nanowire 16. This measurement is done at low count rates – here 82 kcps at saturation. The saturation of the count rate for currents above \SI{10.5}{\micro\ampere} indicates unity internal detection efficiency (IDE), which means that any photon absorbed in the nanowire will create an output pulse that can be detected \cite{Baek2011}. The PCR curve was fit to an idealized model of the IDE as a function of current in the nanowire, 
$\label{eq:efficiency}
                \eta=\frac{1}{2}\mathrm{erfc}\left( -\frac{I_\mathrm{nanowire}-I_\mathrm{detect}}{\sigma}\right)$,
where $I_\mathrm{detect}$ is the inflection point and $\sigma$ is a fit parameter describing the slope of the inflection. At low count rates the current in the nanowire is assumed to be equal to the bias current ($I_\mathrm{nanowire}=I_\mathrm{bias}$) meaning the nanowire is not recovering from a previous detection when it absorbs a photon. Figure~\ref{fig:device}b also shows the dark count rate (DCR) of the same nanowire. In this measurement, dark counts are mostly due to ambient photons coupling to the detector, as indicated by the plateau of 0.4 counts/s at bias currents below $I_\mathrm{switch}$. No filtering or special enclosure was used to minimize the dark count rate. Shortly before $I_\mathrm{switch}$, the DCR shows the typical exponential increase. After $I_\mathrm{switch}$ the nanowire enters a regime of relaxation oscillations, then the count rate starts decreasing, indicating latching. We define $I_\mathrm{latch}$ as the nanowire current at which DCR starts to decrease. Figure \ref{fig:device}c shows $I_\mathrm{detect}$ and $I_\mathrm{switch}$ for all nanowires. $I_\mathrm{detect}$ values were similar, indicating that the nanowires are fairly uniform across the array, though the slightly higher variation in switching current indicates varying degree of constriction in the wires. All wires had large plateaus in IDE vs. current. Since the IDE exceeds $90\%$ for $I_\mathrm{bias} > I_\mathrm{detect} + \sigma$, a measure of the plateau is $I_\mathrm{switch}-(I_\mathrm{detect}+\sigma)$, which ranged from \SI{0.7}{\micro\ampere} to \SI{3.9}{\micro\ampere} for the wires in the array, with an average of $\SI{2.8}{\micro\ampere}$. In order to take advantage of this long detection plateau, the nanowires were biased at 95\% of their individual switching currents for all measurements.

\section{Measurement Setup}

The PEACOQ array was measured in two different cryostats at a temperature of $0.9$~K. One cryostat (Setup A), was used to read out all 32 channels simultaneously. Setup A was initially developed as a testbed for the Deep Space Optical Communications project \cite{Biswas2018}. The second cryostat (Setup B) was used to perform low-jitter measurements of a single detector channel. Further details of this setup can be found in \citenum{Korzh2020}. In each case, the device was wire-bonded to a custom PCB board, designed for differential read-out of each of the 32 nanowires, and packaged to accommodate self-aligned fiber coupling \cite{Miller2011}. One side of each nanowire was terminated with a \SI{50}{\ohm} resistor at $0.9$~K for a single-ended readout of each wire. Future work will focus on a fully differential readout.

In Setup A, microcoaxial cables connected the signal from each nanowire to a cryogenic readout board at 40~K. The cryogenic readout board contained a resistive bias-tee and a two-stage cryogenic amplifier (DC-coupled HEMT and SiGe LNA) for each channel. At room temperature, the pulses were further amplified by a third stage of amplifiers to an amplitude of $\approx500$~mV, then converted to time stamps using a custom 128-channel time-to-digital converter (TDC) from Dotfast Consulting. The optical source comprised either a continuous or pulsed laser at 1550~nm, a variable attenuator, polarization controller, and a switch that sent light either to the PEACOQ or to a power meter. An SMF-28 optical fiber was used to couple light to the detector. The face of the optical fiber ferrule had an anti-reflective coating optimized for 1550 nm. The pulsed laser had a pulse width of $0.5$~ps and a repetition rate of 20~MHz. Using a phase-locked loop, the sync signal of the pulsed laser was converted to a 10~MHz signal to synchronize the TDC. The timing jitter of the readout system (TDC and of the phase-locked loop) was measured to be 71~ps~FWHM. 

Setup B had lower noise and superior timing jitter, but could only measure one nanowire a time. In Setup B, the signal from the SNSPD was amplified with a silicon germanium cryogenic amplifier from Cosmic Microwave Technologies at 4~K (LF1-5K). The amplifier had a built-in bias-tee with a \SI{5}{\kilo\ohm} resistor. In parallel with the amplifier, an inductive shunt to ground (\SI{1.1}{\micro\henry}, \SI{50}{\ohm}) was used to provide a path to ground low frequencies. Jitter measurements were performed with a 1550~nm, 1~GHz mode locked laser and a Swabian time tagger. Note that a larger switching current was measured in this setup compared to Setup A (compare Figures \ref{fig:device}b and Figures \ref{fig:device}c). This was likely due to differences in the readout electronics such as lower load impedance and lower noise on the biasing current in Setup B.

\section{Efficiency and Polarization Sensitivity}\label{sec:efficiency}

\begin{figure}[h]
\centering\includegraphics[width=3in]{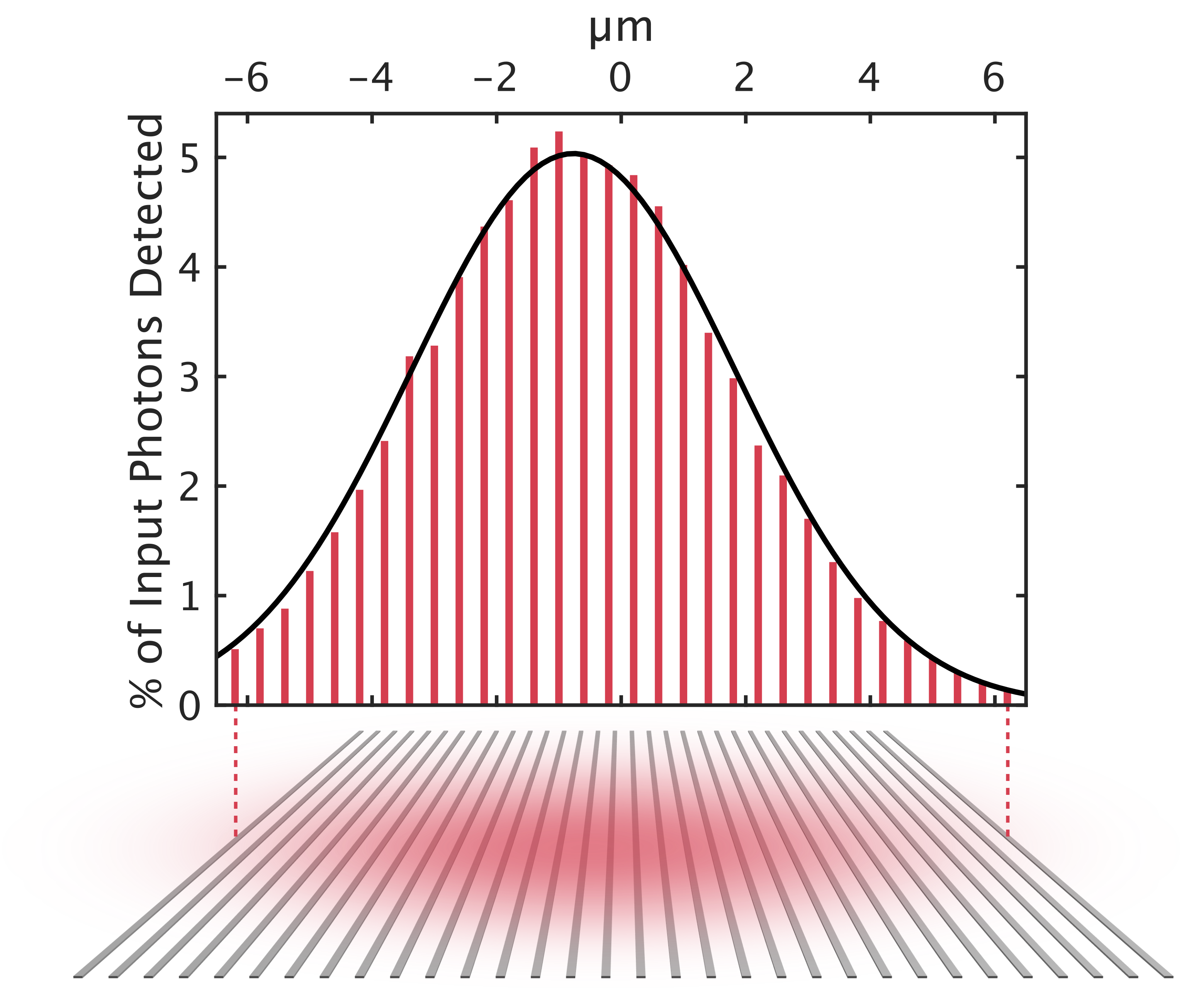}
\caption{\label{fig:efficiency} PEACOQ array sampling SMF-28 optical fiber mode. The height of the red bars represents fraction of input photons detected by each nanowire, the horizontal position of the bars represent the wire position and the width of bars corresponds to the wire width. The optical mode was offset from the center of the array by $-760$ nm. Black curve represents a fit to the Gaussian mode of an SMF-28 optical fiber with a mode field diameter of \SI{10.4}{\micro\meter} at 1550 nm, fit for amplitude and offset. Below the graph is a sketch of the array and optical mode in perspective. The total SDE of the array was 78\%, and the total DCR was 158 counts/s across the array.}
\end{figure}

High SDE in the PEACOQ detector was enabled by good spatial overlap between the optical mode and array in the device plane, and an optical cavity. Figure \ref{fig:efficiency} shows the fraction of input photons detected by each nanowire, measured at a low total array count rate of 7 Mcps. The measured SDE of the array was $78\% \pm 4\%$. The SDE is defined as the fraction of photons incident on the input fiber port of the cryostat that are detected by any wire in the array. The PEACOQ was designed to be polarization insensitive, and we measured no significant difference in SDE while sweeping through input polarization.

With a pitch of 400~nm, the 32-nanowire array covers \SI{12.8}{\micro\meter} contains $98.6\%$ of the power in the optical mode in the case of a perfect alignment. The pitch of 400~nm was found to be the lowest at which there was negligible cross-talk. Cross-talk describes when a detection in one wire leads to an output pulse in a neighboring wire, typically due to thermal effects, leading to double-counting and lower fidelity of detection \cite{Afek2009}. We measured cross-talk contributing less than $0.5\%$ of total counts, which decreased significantly when using a lower bias current (see Supplemental Material). Alignment between the optical mode and the center of the array is also important for achieving high SDE. The mechanical alignment between the optical fiber and the array was robust: the offset differed each time an optical fiber was coupled to the detector, but was always within $\pm$~\SI{1.5}{\micro\meter}, which should lead to a deviation in SDE from the ideally-aligned case of $<1.7\%$. In the self-aligned coupling scheme, the greatest deviations from concentricity are typically in the direction of the slot in the ceramic sleeve. The array wires were therefore oriented along this axis, and the length of the active area was chosen to be longer than its width.

The optical cavity consisted of a gold mirror and a 240 nm layer of SiO\tsub{2} below the nanowire layer, and a 4-layer distributed Bragg reflector composed of SiO\tsub{2}/TiO\tsub{2}/SiO\tsub{2}/TiO\tsub{2} above. The optical stack design was optimized for high SDE at 1550~nm for both polarizations (electric field parallel and perpendicular to the nanowire) using rigorous coupled-wave analysis. The optical absorption bandwidth was $\approx\SI{200}{nm}$ (see Supplemental Material for details). 

\section{Maximum Count Rate}\label{sec:count_rate}

\begin{figure}
\centering\includegraphics[width=3.4in]{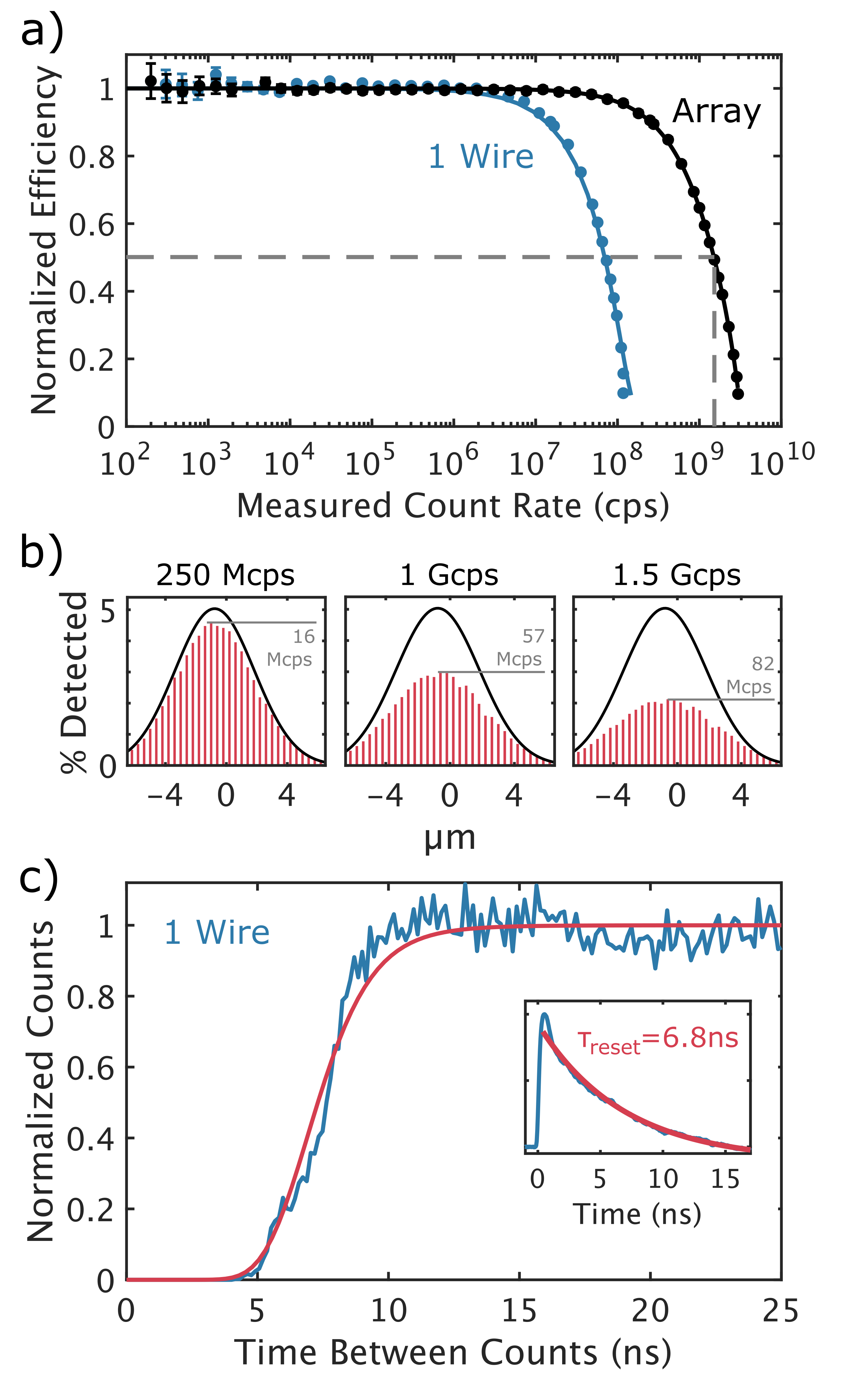}
\caption{\label{fig:MCR} Maximum count rate of array. a) Normalized efficiency of the 32-wire array versus measured count rate (black circles). Error bars show $\sqrt{N_\mathrm{photons}}$ (shot noise) uncertainty. Solid black line is a third order polynomial fit to the data, used to guide the eye and interpolate.  Intersecting dashed lines indicate the 1.5 Gcps MCR of the array (3 dB compression point). Blue circles show the normalized efficiency for nanowire 16. Error bars show shot noise uncertainty. Blue line is a theoretical curve computed using measured parameters and the $\tau_\mathrm{reset}$ value from c). The MCR of this wire was 73 Mcps (3 dB compression). b) SDE of wires across the array at different total array count rates. Black curve is the Gaussian fit from Fig.~\ref{fig:efficiency}. On each panel, the highest single-wire count rate is shown in grey. c) Dead time and reset time. Blue line shows the histogram of counts in nanowire 16 as a function of delay since the last count. For long delays (not shown), the histogram has an exponential decay as predicted by the Poisson statistics of the continuous-wave laser. Red curve is a fit to the theoretical IDE of one nanowire as a function of time since the last detection, computed using measured device parameters, and fit to this data to extract the reset time of this nanowire $\tau_\mathrm{reset}=6.8$ ns. Inset shows a representative pulse from nanowire 16 (blue dots), and an exponential function with a decay time of \SI{6.8}{\nano\second}, fit for amplitude and offset (solid red line).} 
\end{figure}

Detection at high count rates enables applications such as quantum communication at high data rates, time-of-flight mass spectrometry \cite{Casaburi2011}, where faint signals at the are measured in quick succession, and photon time-tagging LIDAR \cite{Barton-Grimley18}, where high dynamic range is important. Figure \ref{fig:MCR}a shows the relative efficiency as a function of the measured count rate for the PEACOQ array and for an individual nanowire. We define the maximum count rate (MCR) as the count rate at which the relative efficiency of detection decreases by 3 dB.

The MCR for the PEACOQ array was measured to be 1.5 Gcps. The MCRs of individual nanowires in the array ranged from 48-73 Mcps. Besides the MCR of individual wires, the array MCR also depends on the profile of the optical mode. The central wires couple more strongly to the optical mode and saturate before the others as shown in Figure \ref{fig:MCR}b. Fig.~\ref{fig:MCR}a also shows a dynamic range of more than 6 orders of magnitude. The largest count rate measured with the array was 3.0 GHz. This was deep in the saturated regime, where the SDE was $8\%$.

The MCR of each nanowire is intrinsically tied to the timescale at which its IDE recovers following a detection. Figure \ref{fig:MCR}c shows a histogram of measured counts in one nanowire as a function of time since the previous count. There were no counts detected during the first $\approx5$~ns after a detection event, which we refer to as the detector dead time. The timescale of efficiency recovery depends on two factors: the reset time of the nanowire, $\tau_\mathrm{reset}$ and the length of its IDE plateau (Fig.~\ref{fig:device}b). We modeled the bias current in the nanowire as a function of time since the previous detection as $I_\mathrm{nanowire}(t)=I_\mathrm{bias}(1-e^{-t/\tau_\mathrm{reset}})$. Using $I_\mathrm{nanowire}(t)$ as the nanowire current in $\eta$, the expression for the nanowire IDE defined in the Overview section gives a theoretical expression for efficiency versus delay $\eta(t)$ which we fit to the histogram in Fig.~\ref{fig:MCR}c at low time delays. We extracted $\tau_\mathrm{reset}=6.8$ ns from this fit. The inset shows that $\tau_\mathrm{reset}=6.8$~ns also visually fits the decay of the pulses as measured on an oscilloscope in Setup B. Fitting the pulse decay to an exponential directly to extract $\tau_\mathrm{reset}$ is less reliable due to filtering and reflections in the readout which distort the pulse shape. 

Using $R_L=\SI{100}{\ohm}$ as the impedance of the readout path, we find $L_k=680$~nH from $\tau_\mathrm{reset}$. Note that $R_L=\SI{100}{\ohm}$ is a sum of the \SI{50}{\ohm} impedance of the amplifier and the \SI{50}{\ohm} resistor on the other end of the nanowire. This setup allowed a lower $\tau_\mathrm{reset} \propto \frac{1}{R_L}$ and higher MCR while avoiding reflections that would result from a mismatch between the amplifier impedance and \SI{50}{\ohm} readout coaxial lines.

Using the method described by Rosenberg et al. \cite{Rosenberg13}, the relative efficiency vs. count rate for one wire in Fig.~\ref{fig:MCR}a (blue curve) was computed from the IDE curve in Fig.~\ref{fig:MCR}b. The agreement between reset time, dead time, and the maximum count rate emphasizes that the limit to the MCR in one wire is well understood (additional discussion in the Supplemental Material). 

The readout chain was optimized for high MCR by using a DC-coupled amplifier as the first amplification stage (or an inductive shunt in Setup B). This configuration provides a well-defined impedance for all frequencies contained within a detection pulse, which improves the MCR by increasing the latching current and therefore the length of the IDE plateau \cite{Kerman2013}. The threshold level of the TDC also influenced the MCR. Any detection event that happens as the current in the nanowire is recovering from a previous detection event will lead to a smaller pulse. To maximize MCR, all pulses must be counted, meaning that the ideal threshold of detection is just above the noise floor of the amplifier chain. We placed the threshold level at $25\%$ of the pulse amplitude for each wire (33\% of pulse amplitude for nanowire 14). See Supplemental Material for more details.

\section{Jitter}\label{sec:jitter}

The timing jitter of a detector characterizes the uncertainty in the measured arrival time of a photon. We measured the jitter by sending in short attenuated laser pulses and collecting a histogram of arrival times, from which we extracted two metrics: the FWHM and the FW1\%M. The FW1\%M metric is important since a significant number of events fall outside of the FWHM for most distributions (24\% for a normal distribution). More importantly, although the intrinsic jitter in SNSPDs can be very small, the instrument response function (IRF) is non-Gaussian~\cite{Korzh2020, Allmaras2019} and thus one needs to characterize the effect of events in the tail of the distribution when considering the impact on applications. In communication protocols based on time bin encoding, it is convenient to set the time-bin to be larger than the FW1\%M value of the IRF in order to reduce errors from counts falling into the wrong time slot below the 1\% level. 

\begin{figure}[h]
\centering\includegraphics[width=3.6in]{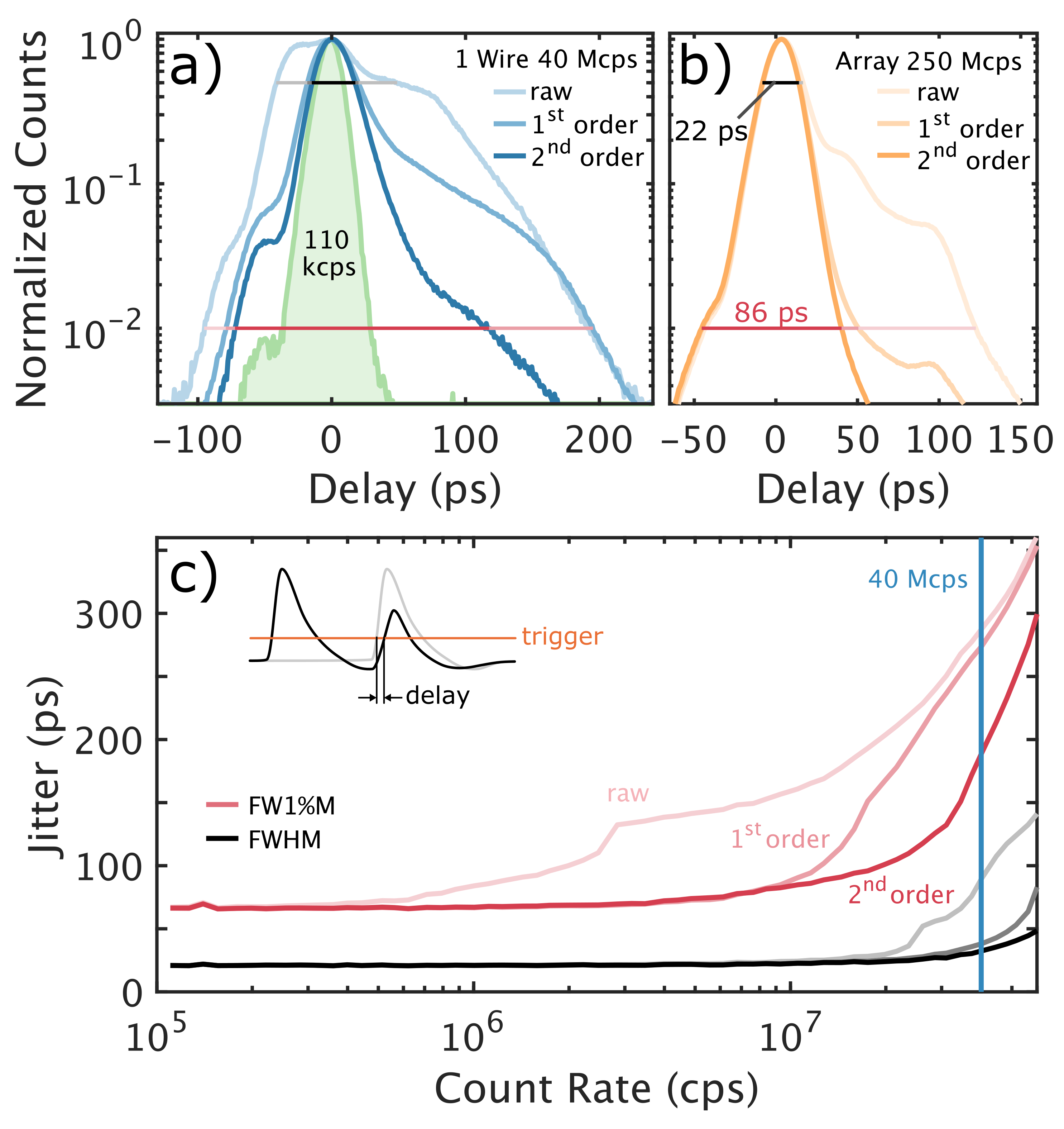}
\caption{\label{fig:jitter} Timing jitter and time-walk correction. a) Jitter histogram for a single wire at a low count rate (filled in green curve) and at 40~Mcps (blue lines). The lightest blue line shows the jitter at 40~Mcps as measured, and the darker (darkest) shade shows a time-walk correction to 1\tsup{st} order (2\tsup{nd} order). The colors of the horizontal lines showing the widths of the histograms have the same meaning as in c). The smaller peak at $\approx - \SI{50}{\pico\second}$ is likely due to two-photon detection events in the nanowire, which have a higher slew rate. b) Estimated jitter for the 32-nanowire array at 250~Mcps. Darker shades indicate time-walk correction, and red and black lines have the same meaning as in c). c) Timing jitter for nanowire 1 as a function of count rate. Lightest red line shows the FWHM of the uncorrected jitter histograms for different count rates. Light gray line shows the corresponding FW1\%M. The darker (darkest) lines show the FWHM and FW1\%M with a 1\tsup{st} (2\tsup{nd}) order time-walk correction. Blue line highlights the jitter at 40~Mcps count rate, represented in detail in a). Inset shows a schematic of time-walk induced jitter.}
\end{figure}

The jitter for nanowire in the array (nanowire 1) was characterized using Setup B. The green curve in Figure \ref{fig:jitter}a shows the histogram of arrival times at low count rates, with a timing jitter of 20.9~ps~FWHM/67.1~ps~FW1\%M. This was limited by the jitter of the readout electronics, and a tighter bound on the intrinsic jitter of the nanowire is 16.8~ps~FWHM/44.7~ps~FW1\%M, which was measured for the same nanowire with a different timing card (see Supplemental Material for details). The low timing jitter for individual wires was a result of the low latency of NbN, the large plateau in IDE versus bias current, and the high bias current used ($0.95 \times I_\mathrm{sw}$), all of which contributed to achieve low intrinsic jitter \cite{Korzh2020}. The small length of the wires (\SI{30}{\micro\meter}) led to low longitudinal geometric jitter \cite{Korzh2020}. The noise jitter was minimized by having pulses with amplitude well above the amplifier noise level and choosing the smallest possible kinetic inductance for the nanowire, leading to a high slew rate \cite{Caloz2019}.

At higher count rates, the timing jitter of SNSPD detectors suffers from time-walk: pulses occurring during detector recovery are smaller, which results in a time-delay when using a constant threshold level to convert pulses to time-tags (see inset to Figure \ref{fig:jitter}c)~\cite{Mueller2022}. The pulse shape can also be distorted due to reflections of previous pulses in the readout chain. Both effects lead to increased timing jitter at higher count rates. The dotted lines in Fig.~\ref{fig:jitter}c shows the jitter FWHM and FW1\%M for nanowire 1 as a function of measured count rate. We used a technique for correcting time-walk-induced jitter, which relies on the fact that the pulse height variation as a function of the time since the previous detection is deterministic, and can be calibrated out~\cite{Mueller2022}. The calibration is performed once on a sample data set, after which it can be used to correct time-walk-induced jitter for additional data either in real-time or in post-processing. We refer to the correction process described in Reference \cite{Mueller2022} as a 1\tsup{st} order correction, since it takes into account the effect of one previous detection on the amplitude of the pulse. We also used a 2\tsup{nd} order time-walk correction, which takes into account the effect of the previous two pulses (see Supplemental Material for details). Figure~\ref{fig:jitter}a shows the uncorrected jitter histogram for nanowire 1 at 40~Mcps, and the effect of both the 1\tsup{st} and 2\tsup{nd} order time-walk correction. Figure~\ref{fig:jitter}c shows the effect of time-walk correction on the jitter for count rates from 110~kcps to 60~Mcps. The 2\tsup{nd} order correction was found to improve the jitter significantly more than the 1\tsup{st} order correction for rates above 10~Mcps.

Using Setup A, the timing jitter of the PEACOQ array was measured to be 84.3~ps~FWHM/202.3~ps~FW1\%M, limited by the jitter of that readout system. To estimate the timing jitter expected for the PEACOQ with an improved measurement setup (i.e.~a multi-channel TDC with lower jitter and a new readout chain with amplification at the 4~K stage), we estimated the array jitter based on the measurements for one nanowire in Setup B. We used the known distribution of count rates across the nanowires for a given array count rate (as in Fig.~\ref{fig:MCR}b) and the jitter histogram for a single nanowire at each count rate (as in Fig~\ref{fig:jitter}a) to construct a simulated histogram of arrival times in the array. Figure \ref{fig:jitter}b shows the estimated jitter histograms for the array at 250~Mcps. Without any correction, there was a large tail in the distribution. The tail was effectively reduced with a 2\tsup{nd} order time-walk correction, leading to a jitter of 22~ps~FWHM/86~ps~FW1\%M, a significant improvement for FW1\%M in particular. 

With a FW1\%M below 100 ps at 250~Mcps, the PEACOQ detector could be used in quantum communication protocols operating at 10 GHz with a relatively high mean photon number per pulse of $\langle n \rangle=0.025$, arriving at the detector. To put this performance into perspective for quantum communication, one can estimate the expected secret key rate of a quantum key distribution (QKD) protocol~\cite{Boaron2018}, where the typical mean photon number per qubit is on the order of 0.2 at the transmitter. This would permit a fiber link of $\approx$100~km (20~dB of link loss assuming 0.2~dB/km loss in standard telecom fiber) while maintaining 250~Mcps at the receiver. With typical secret key fractions, this would correspond to $>70$~Mbps of secret key generation rate, which is almost a factor of 3 faster than the current state-of-the-art~\cite{Islam2017}. Through the use of two or more PEACOQ detectors, QKD secret key rates above 100~Mbps would be achievable. Entanglement swapping protocols, both space-to-ground and fiber based, would also benefit from the PEACOQ's ability to maintain very low jitter at high count rates, especially for approaches based on continuous-wave sources which relieve stringent requirements on source synchronization~\cite{Samara2021}.

Table \ref{tab:jitter} summarizes the timing jitter estimated at three count rates using a 2\tsup{nd} order time-walk correction. At 1~Gcps, the 2\tsup{nd} order time-walk correction was not as effective in reducing jitter (see figure in Supplemental Material). At this count rate, more than half of the nanowires are measuring at 30~Mcps or more, where even the 2\tsup{nd} order correction does not fully compensate count-rate dependent jitter, as seen in Fig.~\ref{fig:jitter}c. It is likely that a correction to 3\tsup{rd} order or higher would improve the jitter further. However, at sufficiently high count rates, the current in the nanowire seldom fully recovers between pulses, leading to a lower effective bias current in the nanowire, which increases the intrinsic jitter of the device \cite{Korzh2020}. The limit where time-walk correction is no longer effective in improving jitter merits further study.

\begin{table}[h]
\centering
\begin{tabular}{|c|c|c|}
\hline
\multicolumn{1}{|c|}{\begin{tabular}[c]{@{}c@{}}Count Rate\\ (Mcps)\end{tabular}} & \multicolumn{1}{|c|}{\begin{tabular}[c]{@{}c@{}}Efficiency\\ (\%)\end{tabular}} & \multicolumn{1}{|c|}{\begin{tabular}[c]{@{}c@{}}Estimated Timing Jitter\\ (ps FWHM)/(ps FW1\%M)\end{tabular}} \\ \hline
7 & 78 & 21 / 66 \\ \hline
250 & 70 & 22 / 86 \\ \hline
1000 & 50 & 46 / 244 \\ \hline
\end{tabular}
\caption{\label{tab:jitter} SDE and estimated timing jitter at different array count rates. Jitter values use a 2\tsup{nd} order time-walk correction.}
\end{table}

\section{Discussion}

Improving the performance of the PEACOQ detector will have to balance the trade-offs between maximum count rate, efficiency and jitter. There are five design parameters to consider: array pitch, nanowire dimensions, kinetic inductance, the superconducting and dielectric materials, and the optical stack.

The pitch of the PEACOQ was chosen as the smallest pitch leading to minimal cross-talk. If the pitch were decreased, the distance between nanowires would be smaller, increasing both thermal and electrical coupling and thereby increasing the likelihood of cross-talk between wires. The array MCR would increase with decreasing pitch, since the center of the optical mode will be shared between a larger number of wires, thereby decreasing saturation in those central wires. The SDE of the array at low count rates would decrease, since the total width of the array is smaller relative to the optical mode (see Supplementary Material for a quantitative analysis). This can be compensated by increasing the number of nanowires in the array, at the cost of increased readout complexity. The loss in efficiency would be mitigated by the increased absorption in the nanowire layer that is expected with lower fill factor.

The width of the PEACOQ nanowires was chosen as a compromise between pulse height, which increases with width, and length of PCR plateau, which decreases with width. If the width of the wires were decreased, the absorption in the nanowire layer would be lower, which could lead to a decrease in array efficiency, though this can be compensated with a higher finesse optical cavity. The consequence of decreasing wire width on MCR is more complicated, since there are several competing effects: (1)  $I_\mathrm{switch} \leq J_\mathrm{critical}\times A_\mathrm{cross-section}$, where $J_\mathrm{critical}$ is the critical current density of the superconductor, would decrease as the cross section of the wire decreases, and since the pulse amplitude is proportional to $I_\mathrm{bias}<I_\mathrm{switch}$, the smaller pulses created by detection events during the recovery of current in the nanowire can become lost in the amplifier noise, which would decrease MCR; (2) as $I_\mathrm{switch}$ decreases, $\tau_\mathrm{reset}$ can be decreased without inducing latching (as explained further on), leading to higher MCR; (3) the plateau in IDE vs. bias current is expected to increase as $I_\mathrm{detect}/I_\mathrm{depairing}$ decreases \cite{Zotova2012} (where $I_\mathrm{depairing}$ is the depairing current in the superconducting film, and an upper bound for $I_\mathrm{switch}$), which would decrease dead time and therefore increase MCR. We found that a decrease in width from 120 nm to 100 nm slightly increased MCR, with an MCR as high as 88~Mcps measured for an individual wire. Decreasing wire width is likely to have a detrimental effect on jitter since noise jitter from the amplifier would have a larger effect as the pulse size becomes smaller. Narrower wires would lead to less thermal cross-talk, as the hotspot energy, which scales with the bias current, would be smaller. However, narrower wires are also likely to have a higher rate of constrictions, which is always detrimental to nanowire performance \cite{Kerman2007}. 

Decreasing the thickness of the superconducting film is another way to decrease $A_\mathrm{cross-section}$, and would therefore have effects similar to those described above. However, the relationship between MCR and nanowire thickness is non-linear, because a thinner superconductor also has a lower critical temperature and therefore a lower critical current \cite{Ivry2014}. We found that decreasing the film thickness by a factor of $\approx 1.5$, as estimated from an increase of $1.5\times$ in film resistance at room temperature, reduced the MCR of individual wires by a factor of $\approx2$.

The total kinetic inductance of the PEACOQ nanowires was controlled by changing the profile of the microstrip transmission line. Both MCR and jitter benefit from a lower kinetic inductance. The former because $\tau_\mathrm{reset} \propto L_k$ is reduced, and the latter because the rise time of the pulses $\tau_\mathrm{rise} \propto L_k$ is reduced, which leads to lower noise jitter \cite{Caloz2019}. However, there is a lower bound to how small $\tau_\mathrm{reset}$ can be: if it is so low that the current returns to the wire before the hotspot dissipates, the nanowire enters a permanently resistive state known as latching \cite{Kerman2009}. The time scale of hotspot dissipation depends on the device temperature, the nanowire and substrate materials, and also on the hotspot energy $E\approx\frac{1}{2}I_\mathrm{bias}L_k^2$ \cite{Kerman2009,Annunziata2010}. Latching occurs at bias currents above a threshold $I_\mathrm{latch}$, which decreases with $\tau_\mathrm{reset}$ \cite{Kerman2013,Annunziata2010}. Given a nanowire with switching current $I_\mathrm{switch}$, a good strategy is to decrease $\tau_\mathrm{reset}$ via $L_k$ until $I_\mathrm{latch}$ is just above $I_\mathrm{switch}$. This way, the nanowires can be biased close to $I_\mathrm{switch}$, the optimal bias point for minimizing jitter and maximizing count rate \cite{Kerman2013,Caloz2019}. As shown in Figure \ref{fig:device}b, the nanowires in this work are close to satisfying this condition. For the 100 nm wide wires measured, the kinetic inductance could be decreased to $\approx430$~nH which led to a faster reset time and higher MCR.

An important design consideration is the choice of superconducting and dielectric materials. Of superconductors known to make efficient SNSPDs at 1550 nm (NbN, tungsten silicide, molybdenum silicide, niobium-titanium nitride \cite{Chang2021}), NbN was chosen as a low-latency material \cite{Korzh2020}, which is important for achieving low timing jitter. Also important for jitter is the interface between superconductor and dielectric, which in this case is SiO\tsub{2}. In future studies, lattice mismatch between the two materials could be used to decrease phonon escape and decrease intrinsic jitter \cite{Kozorezov2017,Allmaras2019}.

The PEACOQ nanowires showed a plateau in PCR, indicating that any photon absorbed by a nanowire was converted to an electric pulse. However, the SDE of the PEACOQ was less than unity, limited by the probability of a input photon being absorbed by a nanowire. Known sources of loss include: 2.5\% loss in the optical fiber in the cryostat, absorption in the gold mirror, estimated to be 5.5\%, and 9\% of light being reflected from the stack due to imperfect impedance matching. Using a distributed Bragg reflector instead of a gold mirror could improve efficiency by eliminating the absorption in the gold \cite{Reddy2020}. Additionally, without absorption in the gold, a higher finesse optical cavity could be designed to increase absorption in the nanowire layer.

The performance of the PEACOQ detector depends on the design parameters in complicated ways: there are trade-offs between different metrics, and in some cases changing one parameter can have two opposing effects on the same metric, so a quantitative study is necessary. There is still room to explore the optimization of PEACOQ-style detectors for specific applications, whether a higher count rate, higher efficiency or lower jitter is desired. 

\section{Conclusion}

We have demonstrated a detector that measures single telecommunication-wavelength photons at high count rates with high SDE and low timing jitter. The 32-nanowire array measured photons from a directly-coupled SMF-28 optical fiber at up to 250~Mcps with an SDE $\geq$~70\%. We measured jitter as low as 44.7~FW1\%M for an individual channel, and we estimated that the timing jitter of the array could be below 100~ps FW1\%M at count rates up to 250~Mcps with an improved readout system. The maximum count rate of the PEACOQ was 1.5~Gcps at 3 dB compression, and the highest measured count rate was 3~Gcps at an SDE of 8\%. This is a factor of two improvement over the highest count rate we are aware of in an SNSPD array \cite{Zhang2019}. Low timing jitter at high count rates was made possible by an improved time-walk correction scheme. Future work should focus on implementing a differential readout architecture, which, together with a lower-jitter multi-channel TDC, would take full advantage of the PEACOQ's low-jitter design.

\section{Acknowledgements}

The research was performed at the Jet Propulsion Laboratory, California Institute of Technology, under contract with NASA. We thank Ryan Lannom for the photograph of the PEACOQ. We thank Gregor Taylor for his helpful comments on the manuscript. We acknowledge funding from NASA SCaN and DARPA DSO, through the DETECT and Invisible Headlights programs. 

\bibliography{peacoq_bibliography}

\providecommand{\noopsort}[1]{}\providecommand{\singleletter}[1]{#1}%
\begin{thebibliography}{10}

\bibitem{Verma2021}
V.~B. Verma, B.~Korzh, A.~B. Walter, A.~E. Lita, R.~M. Briggs, M.~Colangelo,
  Y.~Zhai, E.~E. Wollman, A.~D. Beyer, J.~P. Allmaras, H.~Vora, D.~Zhu,
  E.~Schmidt, A.~G. Kozorezov, K.~K. Berggren, R.~P. Mirin, S.~W. Nam, and
  M.~D. Shaw, ``{Single-photon detection in the mid-infrared up to 10 \textmu m
  wavelength using tungsten silicide superconducting nanowire detectors},''
  {\em APL Photonics}, vol.~6, no.~5, p.~056101, 2021.

\bibitem{Colangelo2022}
M.~Colangelo, A.~B. Walter, B.~A. Korzh, E.~Schmidt, B.~Bumble, A.~E. Lita,
  A.~D. Beyer, J.~P. Allmaras, R.~M. Briggs, A.~G. Kozorezov, E.~E. Wollman,
  M.~D. Shaw, and K.~K. Berggren, ``Large-area superconducting nanowire
  single-photon detectors for operation at wavelengths up to{
  \SI{7.4}{\micro\meter}},'' {\em Nano Letters}, vol.~22, no.~14,
  pp.~5667--5673, 2022.
\newblock PMID: 35848767.

\bibitem{Wollman2017}
E.~E. Wollman, V.~B. Verma, A.~D. Beyer, R.~M. Briggs, B.~Korzh, J.~P.
  Allmaras, F.~Marsili, A.~E. Lita, R.~P. Mirin, S.~W. Nam, and M.~D. Shaw,
  ``{UV} superconducting nanowire single-photon detectors with high efficiency,
  low noise, and {4 K} operating temperature,'' {\em Opt. Express}, vol.~25,
  pp.~26792--26801, Oct 2017.

\bibitem{Reddy2020}
D.~V. Reddy, R.~R. Nerem, S.~W. Nam, R.~P. Mirin, and V.~B. Verma,
  ``Superconducting nanowire single-photon detectors with 98\% system detection
  efficiency at 1550 nm,'' {\em Optica}, vol.~7, pp.~1649--1653, Dec 2020.

\bibitem{Chang2021}
J.~Chang, J.~W.~N. Los, J.~O. Tenorio-Pearl, N.~Noordzij, R.~Gourgues,
  A.~Guardiani, J.~R. Zichi, S.~F. Pereira, H.~P. Urbach, V.~Zwiller, S.~N.
  Dorenbos, and I.~Esmaeil~Zadeh, ``Detecting telecom single photons with
  {$99.5_{-2.07}^{+0.5}\%$} system detection efficiency and high time
  resolution,'' {\em APL Photonics}, vol.~6, no.~3, p.~036114, 2021.

\bibitem{Chiles2022}
J.~Chiles, I.~Charaev, R.~Lasenby, M.~Baryakhtar, J.~Huang, A.~Roshko,
  G.~Burton, M.~Colangelo, K.~Van~Tilburg, A.~Arvanitaki, S.~W. Nam, and K.~K.
  Berggren, ``New constraints on dark photon dark matter with superconducting
  nanowire detectors in an optical haloscope,'' {\em Phys. Rev. Lett.},
  vol.~128, p.~231802, Jun 2022.

\bibitem{Korzh2020}
B.~Korzh, Q.-Y. Zhao, J.~P. Allmaras, S.~Frasca, T.~M. Autry, E.~A. Bersin,
  A.~D. Beyer, R.~M. Briggs, B.~Bumble, M.~Colangelo, G.~M. Crouch, A.~E. Dane,
  T.~Gerrits, A.~E. Lita, F.~Marsili, G.~Moody, C.~Pe{\~n}a, E.~Ramirez, J.~D.
  Rezac, N.~Sinclair, M.~J. Stevens, A.~E. Velasco, V.~B. Verma, E.~E. Wollman,
  S.~Xie, D.~Zhu, P.~D. Hale, M.~Spiropulu, K.~L. Silverman, R.~P. Mirin, S.~W.
  Nam, A.~G. Kozorezov, M.~D. Shaw, and K.~K. Berggren, ``Demonstration of
  sub-3 ps temporal resolution with a superconducting nanowire single-photon
  detector,'' {\em Nature Photonics}, vol.~14, no.~4, pp.~250--255, 2020.

\bibitem{Zhang2019}
W.~Zhang, J.~Huang, C.~Zhang, L.~You, C.~Lv, L.~Zhang, H.~Li, Z.~Wang, and
  X.~Xie, ``A 16-pixel interleaved superconducting nanowire single-photon
  detector array with a maximum count rate exceeding {1.5 GHz},'' {\em IEEE
  Transactions on Applied Superconductivity}, vol.~29, no.~5, pp.~1--4, 2019.

\bibitem{Giustina2015}
M.~Giustina, M.~A.~M. Versteegh, S.~Wengerowsky, J.~Handsteiner, A.~Hochrainer,
  K.~Phelan, F.~Steinlechner, J.~Kofler, J.-A. Larsson, C.~Abell\'an, W.~Amaya,
  V.~Pruneri, M.~W. Mitchell, J.~Beyer, T.~Gerrits, A.~E. Lita, L.~K. Shalm,
  S.~W. Nam, T.~Scheidl, R.~Ursin, B.~Wittmann, and A.~Zeilinger,
  ``Significant-loophole-free test of {Bell's} theorem with entangled
  photons,'' {\em Phys. Rev. Lett.}, vol.~115, p.~250401, Dec 2015.

\bibitem{Zhong2020}
H.-S. Zhong, H.~Wang, Y.-H. Deng, M.-C. Chen, L.-C. Peng, Y.-H. Luo, J.~Qin,
  D.~Wu, X.~Ding, Y.~Hu, P.~Hu, X.-Y. Yang, W.-J. Zhang, H.~Li, Y.~Li,
  X.~Jiang, L.~Gan, G.~Yang, L.~You, Z.~Wang, L.~Li, N.-L. Liu, C.-Y. Lu, and
  J.-W. Pan, ``Quantum computational advantage using photons,'' {\em Science},
  vol.~370, no.~6523, pp.~1460--1463, 2020.

\bibitem{Boaron2018PRL}
A.~Boaron, G.~Boso, D.~Rusca, C.~Vulliez, C.~Autebert, M.~Caloz, M.~Perrenoud,
  G.~Gras, F.~Bussi\`eres, M.-J. Li, D.~Nolan, A.~Martin, and H.~Zbinden,
  ``Secure quantum key distribution over 421 km of optical fiber,'' {\em Phys.
  Rev. Lett.}, vol.~121, p.~190502, Nov 2018.

\bibitem{Grein2015}
M.~E. Grein, A.~J. Kerman, E.~A. Dauler, M.~M. Willis, B.~Romkey, R.~J. Molnar,
  B.~S. Robinson, D.~V. Murphy, and D.~M. Boroson, ``{An optical receiver for
  the Lunar Laser Communication Demonstration based on photon-counting
  superconducting nanowires},'' in {\em Advanced Photon Counting Techniques IX}
  (M.~A. Itzler and J.~C. Campbell, eds.), vol.~9492, p.~949208, International
  Society for Optics and Photonics, SPIE, 2015.

\bibitem{Esmaeil2021}
I.~Esmaeil~Zadeh, J.~Chang, J.~W.~N. Los, S.~Gyger, A.~W. Elshaari,
  S.~Steinhauer, S.~N. Dorenbos, and V.~Zwiller, ``Superconducting nanowire
  single-photon detectors: A perspective on evolution, state-of-the-art, future
  developments, and applications,'' {\em Applied Physics Letters}, vol.~118,
  no.~19, p.~190502, 2021.

\bibitem{Takesue2007}
H.~Takesue, S.~W. Nam, Q.~Zhang, R.~H. Hadfield, T.~Honjo, K.~Tamaki, and
  Y.~Yamamoto, ``Quantum key distribution over a {40-dB} channel loss using
  superconducting single-photon detectors,'' {\em Nat. Photon.}, vol.~1,
  p.~343–348, 2007.

\bibitem{Wang2019}
X.~Wang, B.~A. Korzh, P.~O. Weigel, D.~J. Nemchick, B.~J. Drouin, W.~Becker,
  Q.~Zhao, D.~Zhu, M.~Colangelo, A.~Dane, K.~Berggren, M.~Shaw, and
  S.~Mookherjea, ``Oscilloscopic capture of greater-than-{100 GHz}, ultra-low
  power optical waveforms enabled by integrated electro-optic devices,'' {\em
  Journal of Lightwave Technology}, vol.~38, p.~166, 2019.

\bibitem{Goltsman2001}
G.~N. Gol'tsman, O.~Okunev, G.~Chulkova, A.~Lipatov, A.~Semenov, K.~Smirnov,
  B.~Voronov, A.~Dzardanov, C.~Williams, and R.~Sobolewski, ``Picosecond
  superconducting single-photon optical detector,'' {\em Applied Physics
  Letters}, vol.~79, no.~6, pp.~705--707, 2001.

\bibitem{Cherednichenko2021}
S.~Cherednichenko, N.~Acharya, E.~Novoselov, and V.~Drakinskiy, ``Low kinetic
  inductance superconducting {MgB\textsubscript{2}} nanowires with a 130 ps
  relaxation time for single-photon detection applications,'' {\em
  Superconductor Science and Technology}, vol.~34, p.~044001, feb 2021.

\bibitem{Vetter2016}
A.~Vetter, S.~Ferrari, P.~Rath, R.~Alaee, O.~Kahl, V.~Kovalyuk, S.~Diewald,
  G.~N. Goltsman, A.~Korneev, C.~Rockstuhl, and W.~H.~P. Pernice,
  ``Cavity-enhanced and ultrafast superconducting single-photon detectors,''
  {\em Nano Letters}, vol.~16, pp.~7085--7092, 11 2016.

\bibitem{Korneev2007}
A.~Korneev, O.~Minaeva, A.~Divochiy, A.~Antipov, N.~Kaurova, V.~Seleznev,
  B.~Voronov, G.~Gol'tsman, D.~Pan, J.~Kitaygorsky, W.~Slysz, and
  R.~Sobolewski, ``{Ultrafast and high quantum efficiency large-area
  superconducting single-photon detectors},'' in {\em Photon Counting
  Applications, Quantum Optics, and Quantum Cryptography} (M.~Dusek, M.~S.
  Hillery, W.~P. Schleich, I.~Prochazka, A.~L. Migdall, and A.~Pauchard, eds.),
  vol.~6583, pp.~165 -- 173, International Society for Optics and Photonics,
  SPIE, 2007.

\bibitem{Perrenoud2021}
M.~Perrenoud, M.~Caloz, E.~Amri, C.~Autebert, C.~Schönenberger, H.~Zbinden,
  and F.~Bussi{\`{e}}res, ``Operation of parallel {SNSPDs} at high detection
  rates,'' {\em Superconductor Science and Technology}, vol.~34, p.~024002, jan
  2021.

\bibitem{Allmaras2017}
J.~P. Allmaras, A.~D. Beyer, R.~M. Briggs, F.~Marsili, M.~D. Shaw, G.~V. Resta,
  J.~A. Stern, V.~B. Verma, R.~P. Mirin, S.~W. Nam, and W.~H. Farr,
  ``Large-area 64-pixel array of {WSi} superconducting nanowire single photon
  detectors,'' in {\em 2017 Conference on Lasers and Electro-Optics (CLEO)},
  pp.~1--2, 2017.

\bibitem{Zhao2017}
Q.-Y. Zhao, D.~Zhu, N.~Calandri, A.~E. Dane, A.~N. McCaughan, F.~Bellei, H.-Z.
  Wang, D.~F. Santavicca, and K.~K. Berggren, ``Single-photon imager based on a
  superconducting nanowire delay line,'' {\em Nature Photonics}, vol.~11,
  no.~4, pp.~247--251, 2017.

\bibitem{Mueller2022}
A.~Mueller, E.~E. Wollman, B.~Korzh, A.~D. Beyer, L.~Narvaez, R.~Rogalin,
  M.~Spiropulu, and M.~D. Shaw, ``Time-walk and jitter correction in snspds at
  high count rates,'' {\em arXiv:2210.01271}, 2022.

\bibitem{Colangelo2021}
M.~Colangelo, B.~Korzh, J.~P. Allmaras, A.~D. Beyer, A.~S. Mueller, R.~M.
  Briggs, B.~Bumble, M.~Runyan, M.~J. Stevens, A.~N. McCaughan, D.~Zhu,
  S.~Smith, W.~Becker, L.~Narváez, J.~C. Bienfang, S.~Frasca, A.~E. Velasco,
  C.~H. Peña, E.~E. Ramirez, A.~B. Walter, E.~Schmidt, E.~E. Wollman,
  M.~Spiropulu, R.~Mirin, S.~W. Nam, K.~K. Berggren, and M.~D. Shaw,
  ``Impedance-matched differential superconducting nanowire detectors,'' {\em
  arXiv:2108.07962}, 2021.

\bibitem{Kerman2009}
A.~J. Kerman, J.~K.~W. Yang, R.~J. Molnar, E.~A. Dauler, and K.~K. Berggren,
  ``Electrothermal feedback in superconducting nanowire single-photon
  detectors,'' {\em Phys. Rev. B}, vol.~79, p.~100509, Mar 2009.

\bibitem{Miller2011}
A.~J. Miller, A.~E. Lita, B.~Calkins, I.~Vayshenker, S.~M. Gruber, and S.~W.
  Nam, ``Compact cryogenic self-aligning fiber-to-detector coupling with losses
  below one percent,'' {\em Opt. Express}, vol.~19, pp.~9102--9110, May 2011.

\bibitem{Baek2011}
B.~Baek, A.~E. Lita, V.~Verma, and S.~W. Nam, ``Superconducting
  {a-W$_x$Si$_{1-x}$} nanowire single-photon detector with saturated internal
  quantum efficiency from visible to 1850 nm,'' {\em Applied Physics Letters},
  vol.~98, no.~25, p.~251105, 2011.

\bibitem{Biswas2018}
A.~Biswas, M.~Srinivasan, S.~Piazzolla, and D.~Hoppe, ``{Deep space optical
  communications},'' in {\em Free-Space Laser Communication and Atmospheric
  Propagation XXX} (H.~Hemmati and D.~M. Boroson, eds.), vol.~10524,
  p.~105240U, International Society for Optics and Photonics, SPIE, 2018.

\bibitem{Afek2009}
I.~Afek, A.~Natan, O.~Ambar, and Y.~Silberberg, ``Quantum state measurements
  using multipixel photon detectors,'' {\em Phys. Rev. A}, vol.~79, p.~043830,
  Apr 2009.

\bibitem{Casaburi2011}
A.~Casaburi, M.~Ejrnaes, N.~Zen, M.~Ohkubo, S.~Pagano, and R.~Cristiano,
  ``Thicker, more efficient superconducting strip-line detectors for high
  throughput macromolecules analysis,'' {\em Applied Physics Letters}, vol.~98,
  no.~2, p.~023702, 2011.

\bibitem{Barton-Grimley18}
R.~A. Barton-Grimley, R.~A. Stillwell, and J.~P. Thayer, ``High resolution
  photon time-tagging {LIDAR} for atmospheric point cloud generation,'' {\em
  Opt. Express}, vol.~26, pp.~26030--26044, Oct 2018.

\bibitem{Rosenberg13}
D.~Rosenberg, A.~J. Kerman, R.~J. Molnar, and E.~A. Dauler, ``High-speed and
  high-efficiency superconducting nanowire single photon detector array,'' {\em
  Opt. Express}, vol.~21, pp.~1440--1447, Jan 2013.

\bibitem{Kerman2013}
A.~J. Kerman, D.~Rosenberg, R.~J. Molnar, and E.~A. Dauler, ``Readout of
  superconducting nanowire single-photon detectors at high count rates,'' {\em
  Journal of Applied Physics}, vol.~113, no.~14, p.~144511, 2013.

\bibitem{Allmaras2019}
J.~Allmaras, A.~Kozorezov, B.~Korzh, K.~Berggren, and M.~Shaw, ``Intrinsic
  timing jitter and latency in superconducting nanowire single-photon
  detectors,'' {\em Phys. Rev. Applied}, vol.~11, p.~034062, Mar 2019.

\bibitem{Caloz2019}
M.~Caloz, B.~Korzh, E.~Ramirez, C.~Sch\"onenberger, R.~J. Warburton,
  H.~Zbinden, M.~D. Shaw, and F.~Bussi\`eres, ``Intrinsically-limited timing
  jitter in molybdenum silicide superconducting nanowire single-photon
  detectors,'' {\em Journal of Applied Physics}, vol.~126, no.~16, p.~164501,
  2019.

\bibitem{Boaron2018}
A.~Boaron, B.~Korzh, R.~Houlmann, G.~Boso, D.~Rusca, S.~Gray, M.~J. Li,
  D.~Nolan, A.~Martin, and H.~Zbinden, ``Simple 2.5 ghz time-bin quantum key
  distribution,'' {\em Applied Physics Letters}, vol.~112, p.~171108, 2018.

\bibitem{Islam2017}
N.~T. Islam, C.~C.~W. Lim, C.~Cahall, J.~Kim, and D.~J. Gauthier, ``Provably
  secure and high-rate quantum key distribution with time-bin qudits,'' {\em
  Science Advances}, vol.~3, pp.~1--7, 2017.

\bibitem{Samara2021}
F.~Samara, N.~Maring, A.~Martin, A.~S. Raja, T.~J. Kippenberg, H.~Zbinden, and
  R.~Thew, ``Entanglement swapping between independent and asynchronous
  integrated photon-pair sources,'' {\em Quantum Science and Technology},
  vol.~6, p.~045024, sep 2021.

\bibitem{Zotova2012}
A.~N. Zotova and D.~Y. Vodolazov, ``Photon detection by current-carrying
  superconducting film: A time-dependent {Ginzburg-Landau} approach,'' {\em
  Phys. Rev. B}, vol.~85, p.~024509, Jan 2012.

\bibitem{Kerman2007}
A.~J. Kerman, E.~A. Dauler, J.~K.~W. Yang, K.~M. Rosfjord, V.~Anant, K.~K.
  Berggren, G.~N. {Gol'tsman}, and B.~M. Voronov, ``Constriction-limited
  detection efficiency of superconducting nanowire single-photon detectors,''
  {\em Applied Physics Letters}, vol.~90, no.~10, p.~101110, 2007.

\bibitem{Ivry2014}
Y.~Ivry, C.-S. Kim, A.~E. Dane, D.~De~Fazio, A.~N. McCaughan, K.~A. Sunter,
  Q.~Zhao, and K.~K. Berggren, ``Universal scaling of the critical temperature
  for thin films near the superconducting-to-insulating transition,'' {\em
  Phys. Rev. B}, vol.~90, p.~214515, Dec 2014.

\bibitem{Annunziata2010}
A.~J. Annunziata, O.~Quaranta, D.~F. Santavicca, A.~Casaburi, L.~Frunzio,
  M.~Ejrnaes, M.~J. Rooks, R.~Cristiano, S.~Pagano, A.~Frydman, and D.~E.
  Prober, ``Reset dynamics and latching in niobium superconducting nanowire
  single-photon detectors,'' {\em Journal of Applied Physics}, vol.~108, no.~8,
  p.~084507, 2010.

\bibitem{Kozorezov2017}
A.~G. Kozorezov, C.~Lambert, F.~Marsili, M.~J. Stevens, V.~B. Verma, J.~P.
  Allmaras, M.~D. Shaw, R.~P. Mirin, and S.~W. Nam, ``Fano fluctuations in
  superconducting-nanowire single-photon detectors,'' {\em Phys. Rev. B},
  vol.~96, p.~054507, Aug 2017.

\end{thebibliography}


\end{document}